\documentclass{sig-alternate-05-2015}
\usepackage[utf8]{inputenc}
\usepackage{amsmath}
\usepackage{multirow}
\usepackage{amsfonts}
\usepackage{amssymb}
\usepackage{graphicx}
\usepackage{adjustbox}
\usepackage{color}
\usepackage{multirow}
\usepackage[table]{xcolor}
\usepackage{float}

\begin{document}

\CopyrightYear{2016}
\setcopyright{acmcopyright}
\conferenceinfo{CIKM'16 ,}{October 24-28, 2016, Indianapolis, IN, USA}
\isbn{978-1-4503-4073-1/16/10}\acmPrice{\$15.00}
\doi{http://dx.doi.org/10.1145/2983323.2983882}
\clubpenalty=10000
\widowpenalty = 10000

\title{Ensemble Learned Vaccination Uptake Prediction using Web Search Queries}

\numberofauthors{3} 
\author{
\alignauthor
Niels Dalum Hansen\\
       \affaddr{University of Copenhagen \\ IBM Denmark}\\
       \email{nhansen@di.ku.dk}
\alignauthor
Christina Lioma\\
       \affaddr{University of Copenhagen}\\
       \email{c.lioma@di.ku.dk}
\alignauthor
Kåre Mølbak\\
       \affaddr{Statens Serum Institut}\\
       \email{KRM@ssi.dk}       
}

\maketitle

\begin{abstract}

%Models for predicting health events, such as influenza activity, has successfully been improved by adding web data, such as query frequencies. We show that the prediction of vaccination uptake can also be improved by adding web data. In addition we extend current methods for combining health data with web data on two parts: (i) query generation and (ii) predicting based on query frequencies. We evaluate our method on 13 vaccines in the Danish children vaccination program and show improved prediction quality for 10/13 vaccines using our method compared to forecasting based on only vaccination uptake data or web data.
We present a method that uses ensemble learning to combine clinical and web-mined time-series data in order to predict future vaccination uptake. 
The clinical data is official vaccination registries, and the web data is query frequencies collected from Google Trends. 
Experiments with official vaccine records show that our method predicts vaccination uptake effectively (4.7 Root Mean Squared Error). Whereas performance is best when combining clinical and web data, using solely web data yields comparative performance. 
%), which shows the potential for significant improvements with optimising settings. We further show that predictions based only on web data are close to predictions combining clinical and web data. This indicates that web data can be a complementary data source of vaccination uptake in countries that do not have a national vaccination registry. 
To our knowledge, this is the first study to predict vaccination uptake using web data (with and without clinical data).

\end{abstract}

% A category with the (minimum) three required fields
%\category{H.3.3}{Information Search and Retrieval}{}

%\keywords{}

\section{Introduction and Related Work}
Predicting public health events, e.g.~how many people may get vaccinated in the near future, %is an area of research interest.
%, partially because of the increasing availability of health related data. 
%Computationally, predicting an event given large amounts of heterogeneous time-series data is a challenge in itself. 
%Medically, being able to make such predictions 
can reduce the reaction time of public health professionals, resulting in more efficient services and 
improved public health. 
Traditionally, public health event prediction relied on \textit{clinical data} (e.g.~microbiological results or patient registries) 
that was collected from designated bodies. In the last decade however, non-clinical \textit{web data} (e.g.~search engine queries or microblog messages), 
has been shown useful to the task of predicting public health events. Clinical and web data are complementary sources of evidence: 
Whereas clinical data contributes expert and curated information to the prediction, web data contributes near real-time information on a large scale about e.g.~symptoms or health concerns that may go undetected or unreported by the official clinical channels. %Consequently, web data has great potential in e.g. reducing the reaction time of public health professionals. 

We present a method for predicting vaccination uptake by combining clinical and web data using ensemble learning. 
Combining such clinical and web search data for vaccination uptake prediction is novel. 
So far, research on vaccination uptake has focused on the effect of physician recommendations on vaccination 
uptake \cite{gargano2013impact}; how combined sources of information (e.g. physician, television, friends) 
influence people's decisions about vaccination \cite{gargano2015influence}; and the effects of media coverage on vaccination uptake with respect to influenza vaccination \cite{ma2006influenza}, HPV vaccination \cite{kelly2009hpv}, 
and MMR vaccination \cite{%mason2000impact, 
	smith2008media}. To our knowledge, our study is the first to predict vaccination uptake using web data (with and without clinical data). %Predicting vaccination uptake based only on web data is interesting in itself, as some countries do not have vaccination registries covering the whole population, and web data could in those places constitute a timely counterpart to the survey-based reports of vaccination uptake.

%A connection between web data and vaccination uptake has currently not be studied. 
%In addition prediction of vaccination uptake is currently not performed. 
%One reason for this could be the fact that vaccination uptake is traditionally computed based on birth cohorts. Using this metric for vaccination uptake makes weekly or monthly predictions difficult.

%We propose combining web query logs with official health records on vaccination uptake to predict future vaccinate uptake. 
Web and/or clinical data have been used before for other types of health event predictions, 
e.g.~influenza activity 
\cite{ginsberg2009detecting, mciver2014wikipedia, paul2014twitter, polgreen2008using, santillana2014using, yuan2013monitoring}, dengue fever \cite{chan2011using} and cholera \cite{chunara2012social}.  How the different types of data should be handled has evolved from using a unified model for  both web and clinical data \cite{lazer2014parable, yang2015accurate}, to using ensemble methods that model separately clinical and web data and then combine the outputs \cite{santillana2015combining}.
%In their work they did not take advantage of the fact that applying forecasting methods to the health data should improve prediction accuracy. 
When web search query frequencies are used for prediction \cite{santillana2015combining, santillana2014using, yang2015accurate}, 
a single linear model is used to combine the query frequencies into a prediction. 
%However, whereas there is often a limited amount of clinical data samples for the health event, the number of queries tends to be high, 
%leading to situations with more explanatory 
%variables than training data points as in \cite{yang2015accurate}. The predictions tend therefore to be noisy and unstable. 
Methods using query frequencies select queries either by (i) timely correlation between query search frequency and 
the health event \cite{ginsberg2009detecting, santillana2015combining, santillana2014using, yang2015accurate}, or by (ii) expert selection of 
queries \cite{chan2011using, polgreen2008using, yuan2013monitoring}. 
Both approaches have disadvantages. Approach (i) relies on calculating the correlation between the health event time-series and all queries, which is computationally expensive. It also assumes that historic correlation equals predictive power in the future, which may not always be the case. Approach (ii) relies on human experts, which is costly and does not scale well. In this work we propose a third approach: We select queries based on web descriptions of the health event, in our case of the vaccine in question, and we use an ensemble learning approach, specifically stacking, to predict vaccination uptake. 

\section{Ensemble Learning Prediction}\label{s:predicting_health}

%\subsection{Automatic query generation}\label{s:query_generation}

%\subsection{Prediction vaccination uptake}

Vaccination uptake prediction with time-series data can be formulated as: 
%\begin{equation}
%\label{eq:forecast}
$\hat{E}(t) \approx E(t-1)$, 
%\end{equation}
%\noindent 
where $\hat{E}(t)$ is the predicted vaccination uptake at time $t$, and $E(t-1)$ is the observed vaccination uptake at time $t-1$. 
%We will in Section \ref{s:forecasting_clinical} show how this prediction can be improved by using more historic information about the vaccination uptake. 
%In Section \ref{s:forecasting_web} we will show how web data, in the form of query frequency time-series, can be used to predict vaccination uptake. 
%\subsection{Ensemble learning}
%\label{sec:combining_clinical_and_web}
%We will describe the general approach that we propose, then we will show how it can be applied to already published methods. Finally we will use our approach on a novel data set and show that improvements in prediction quality can be obtained on the predictions on all the children vaccines.
%\subsection{Ensemble approach to forecasting with online data}
% stacking: I think that this can be categorized as stacking, where we use as a combiner a linear model. An example of a stacking algorithm is the winner of the netflix competition "The BellKor Solution to the Netﬂix Grand Prize".
We compute $\hat{E}(t)$ using ensemble learning by combining separate predictions on vaccination uptake based on clinical 
and web data into one prediction. 
Ensemble learning combines predictions from an ensemble of level-0 models into one prediction using a level-1 meta model. 
We use an ensemble method called \textit{stacking}. %, where a level-1 model is trained to combine the predictions of other level-0 models \cite{Witten:2011:DMP:1972514}. 
First, all level-0 models are trained. 
Then, a level-1 model is trained to make a final prediction using all the predictions of the level-0 models as input. 
We experiment with three different types of level-1 models: a linear model, support vector regression (SVR) with a linear kernel, 
and SVR  with a Gaussian kernel. Both our clinical and web data are time-series, i.e.~each data point has a temporal reference.

\subsection{Level-1 models}
\noindent\textbf{Stacking with linear model.} We define a linear model with two explanatory variables:
%\begin{equation}
%\label{eq:ensemble}
$\hat{E}(t) = \mu +  \beta_1 \hat{E}_c(t) + \beta_2 \hat{E}_w(t)$,
%\end{equation}
where $\hat{E}_c(t)$ is the prediction based on clinical data at time $t$,   
$\hat{E}_w(t)$ is the prediction based on web data at time $t$, and $\mu$, $\beta_1$ and $\beta_2$ denote the coefficients that need to be optimized. We use ordinary least squares to find the coefficients that minimize:
%\begin{equation}
$\underset{\mu, \beta_1, \beta_2}{min} \sum_t \left(E(t) - \mu - \beta_1 \hat{E}_c(t) - \beta_2 \hat{E}_w(t)\right)^2$.
%\end{equation}

\noindent\textbf{Stacking with SVR.} SVR solves the same problem as the linear model presented above, but with the possibility of using kernels to transform the input into another feature space.
In addition $\mu$, $\beta_1$ and $\beta_2$ are selected to minimize the following:
%\begin{equation}
%\begin{aligned}
$\underset{\mu, \beta_1, \beta_2}{min} \sum_t V(E(t) - \mu - \beta_1 \hat{E}_c(t) - \beta_2 \hat{E}_w(t)) + \frac{\lambda}{2} (\mu^2 + \beta_1^2 + \beta_2^2)$,
%\end{aligned}
%\end{equation}
where $\lambda$ is a hyperparameter controlling the penalty for large coefficients, and $V(r)$ is defined as $0$ if $|r| < \epsilon$ and otherwise $|r|-\epsilon$. The parameter $\epsilon$ controls how precise the prediction has to be before it is treated as correct.

We experiment with an SVR with linear kernel and with a Gaussian kernel defined as:
%\begin{equation}
$K(x,x') = \exp(-\gamma ||x-x'||^2)$,
%\end{equation}
where $\gamma$ is a hyperparameter.

\subsection{Level-0 models}
\noindent \textbf{Prediction with clinical data.} As level-0 models we use 
three well-known time-series methods: autoregressive (AR) models \cite{yang2015accurate,lazer2014parable}, ARIMA and Holt Winters (HW). 

AR models estimate $\hat{E}(t)$ as: 
%\begin{equation}
%\label{eq:autoregressive}
$\hat{E}(t) = \mu + \sum^m_{i=1} \beta_i E(t-i)$ 
%\end{equation}
where $m$ is the number of autoregressive terms, $\mu$ is the intercept, and the $\beta$s control the weight that each past observation has on the prediction. AR models assume that future values of $E$ can be predicted by a linear combination of the $m$ most recently observed values of $E$. 
With enough autoregressive terms AR models can handle seasonal changes, but not general upwards or downwards trends.

%\paragraph{ARIMA}

An extension of the AR models are the ARIMA (AutoRegressive Integrated Moving Average) models. In addition to the autoregressive terms, these models also include a moving average, which is a weighted sum of the $q$ most recent forecasting errors. 
Let $m$ denote the number of autoregressive terms and $q$ the number of moving averages; then:
%\begin{equation}\label{eq:arma}
$\hat{E}(t) = \mu + \sum^m_{i=1} \beta_i E(t-i) + \sum^q_{j=1} \phi_j \epsilon_{t-j} + \epsilon_t$,
%\end{equation}
where $\epsilon_t = E(t) - \hat{E}(t)$. To handle trend, the original signal $E$ can be differentiated one or more times \cite{chatfield2013analysis}.%, if differentiation has been used it is necessary to integrate the results of Equation \ref{eq:arma} to obtain the original time series.

%\paragraph{Holt Winters forecasting}

HW forecasting is defined by three recursive equations controlling: level, trend and seasonality. HW can forecast time-series with both trend and seasonal changes. 
Each equation is defined as a weighted sum in which the weight of historic observations decreases exponentially with time. 
HW forecasting with level, trend and seasonality is recursively defined as:

\begin{equation}\label{eq:hw}
\begin{aligned}
\text{level} & \quad &a_t &= \alpha (E(t) - s_{t-l}) + (1-\alpha) (a_{t-1} + b_{t-1})\\
\text{trend} & \quad &b_t &= \beta (a_t - a_{t-1}) + (1-\beta) b_{t-1}\\
\text{seasonality} & \quad& s_t &= \gamma (E(t) - a_t) + (1-\gamma) s_{t-l}\\
\end{aligned}
\end{equation}

\noindent where $l$ is the length of the season and $\alpha$, $\beta$ and $\gamma$ are the smoothing parameters which control the influence of the historic level, trend and seasonality. 
Predictions are made by combining level, trend and seasonality:
%\begin{equation}
$\hat{E}(t) = a_{t-1} + b_{t-1} + s_{t - l + 1}$. 
%\end{equation}

%\paragraph{ETS}

%ETS is an extension of HW forecasting and also uses level, trend and seasonality. While HW is defined as having an additive error, trend and seasonality, 
%ETS allows for both additive and multiplicative error, trend and seasonality. In contrast to HW, the multiplicative terms allow ETS to for example model 
%a time-series with an exponential trend as opposed to only linear. The definition of ETS is similar to Equation \ref{eq:hw}, but each different combination of error, 
%trend and seasonality has individual formulations (see \cite{hyndman2007automatic} for more).

\noindent \textbf{Prediction with web data.} As level-0 models we use a linear model, bagging and weighted majority. 
%All three methods assume that the web data is available without any delay. This means that we can predict the health event $E$ at time $t$ using web data available at time $t$. ** NDH: we are actually using data available at the same time as we predict, so maybe it should be denoted 'now cast'? **
Our web data consists of time-stamped query frequencies (described in Section \ref{s:eval}).

\noindent\textit{Linear model.} Given a collection of $n$ query frequency time-series, denoted $Q$, we define a simple linear model as:
%\begin{equation} \label{eq:linear}
$\hat{E}(t) = \mu + \sum_{i=1}^n \alpha_i Q_i(t)$,
%\end{equation}
where $\mu$ and $\alpha$ are coefficients to be estimated. Such a model can be fitted using any of 
several methods, %\cite[Chapter 3]{bishop2006pattern}, 
the most common being ordinary least squares. 
Another approach is to use LASSO regularization which is commonly used for making predictions using query frequencies \cite{santillana2015combining, santillana2014using, yang2015accurate}. This approach adds an additional constraint to the optimization, namely that the sum of the coefficients should also be minimized. 
The weight of this sum is controlled by the hyperparameter $\lambda$. This approach can be used to avoid overfitting and to reduce the coefficients of non-informative 
features to zero and thereby induce a sparse model. 
%%\cite[Chapter 3]{bishop2006pattern}. 
This is a useful property in this context because the collection of queries might 
contain non-informative terms.

\noindent\textit{Bagging.} With bagging, we consider the average of the predictions made on subsets of the training data. This helps to reduce variance and overfitting. We generate subsets of the training data by uniformly sampling with replacement $n$ datasets of size $m$. For each dataset a linear model, as defined above, is fitted using LASSO regularization, 
where the parameter $\lambda$ is found using 3-fold cross-validation. The prediction of the ensemble is the average of the $n$ predictions.

\noindent\textit{Weighted Majority.} We extend the bagging approach to a boosting approach using a weighted majority (WM) algorithm \cite{littlestone1994weighted}. 
The WM algorithm works by combining predictions from a collection of models using a weighted average. Each model is associated with its own weight related to its 
previous predictive performance. If the overall prediction is wrong by a constant $\epsilon$, the weights are updated. The updating works as follows: 
if the individual prediction of a model has an error $> \epsilon$, a new weight is calculated as $w_i = w_i \exp(-\eta)$, where $w_i$ is the weight for 
model $i$ and $\eta$ is a hyperparameter controlling the penalty for making wrong predictions. Our collection of models is identical to the models used for the 
bagging approach described above.

%\begin{verbatim}
%weighted_majority(eta, epsilon, start, end, 
%                  experts, target):
%                  
%    w = [1/length(experts)] * length(experts)
%    
%    for t in start:end {
%        p = predict(experts, t)
%        y_hat = weighted_mean(p, w)
%        y = target[t]
%        if (abs(y-y_hat) > epsilon) {
%            w_new = []
%            for i in 1:length(experts) {
%                if abs(p[i]-y) > epsilon {
%                    w_new[i] = w[i] * exp(-eta)
%                }
%            }
%            w = w_new / sum(w_new)
%        }
%    }
%\end{verbatim}

\begin{table*}
\scalebox{0.64}{
	\begin{tabular}{ll}
		  \textbf{Vaccine} & \textbf{Terms in Danish (English)}\\ 
		  \hline
		   MMR &levende (alive), mæslinger (measles), vaccine, vaccinen (the vaccine), udbrud (outbreak), alvorlige (serious),  fåresyge (mumps), måneders (months), undersøgelser (examinations)\\
		   &beskyttelse (protection), voksne (adults), gravid (pregnant), kombineret (combined), dosis, hunde (dogs), alderen (the age), hjernebetændelse (inflammation of the brain)\\
		   &lungebetændelse (pneumonia), gives (is given), mfr (mmr), røde (red)\\
		   DiTeKiPol & mæslinger (measles), vaccinen (the vaccine), alvorlige (serious), beskyttelse (protection), kombineret (combined), vaccination, indeholder (contains), type, beskytter (protects)\\ 		&sygdomme (illness), meningitis, forårsaget (caused), dræbte (killed), b, kighoste (whooping cough), vare (lasts), polio, difteri (diphtheria), mindst (least), stivkrampe (tetanus)                  \\
		   PCV & vaccinen (the vaccine), alvorlige (serious), alderen (the age), lungebetændelse (pneumonia), vaccination,  infektioner (infections), sygdomme (illness), forebygger (prevents)\\ 				&meningitis, forårsaget (caused), antal (number), blodforgiftning (blood poisoning)  \\
		   HPV & beskyttelse (protection), gives (is given), vaccination, tilbuddet (the offer), kondylomer (condyloma), doser (doses), kønsvorter (genital warts), tilbydes (is offered), piger (girls)\\ 	&livmoderhalskræft (cervical cancer), forventes (is expected), indeholder (contains), januar (january), langvarig (long term), indført (introduced), tilbud (offer), type, human\\ 
		   &beskytter (protects), effekten (the effect), skyldes (caused by), hpv, pigerne (the girls)\\
	\end{tabular} 
	}
	\caption{Our 58 queries.}
\label{tab:queries}

\end{table*}

\section{Experimental Evaluation} \label{s:eval}

%We will start by describing what data set has been used, then how the experiments have been executed, the evaluation metric, the tuning parameters and finally the results.

\noindent\textbf{Data\footnote{\scriptsize{All our data is freely available at: \url{https://sid.erda.dk/share_redirect/c7j6MdrscL}}}.} We evaluate the effectiveness of our approach in predicting vaccination uptake in Denmark for 
all official children vaccines: DiTeKiPol-1, DiTeKiPol-2, DiTeKiPol-3, DiTeKiPol-4, PCV-1, 
PCV-2, PCV-3, MMR-1, MMR-2(4), MMR-2(12), HPV-1, HPV-2 and HPV-3. We use as clinical data the actual vaccination uptake recorded by the country's official body, the State Serum Institut. Specifically, the vaccination uptake is the total number of vaccines given in a month divided by the number of people expected to be vaccinated that month 
(based on the size of the monthly birth cohorts published by Statistics Denmark).%\cite{dst}).% -- see Table \ref{tab:registrations}. 
%We assume that people follow the vaccination schedule recommended by SSI \cite{ssi_children_vaccines}. 
%The recommended vaccination ages and total number of given 
%vaccines in the time period are shown in . We aggregate the data on a monthly basis. % to comply with the web data. The data regards only Denmark.

We use as web data web search queries that are related to each vaccine. We generate these queries from descriptions of each vaccine in: \textit{www.ssi.dk}, 
%(the homepage for the Danish center for disease control, 
\textit{www.patienthåndbogen.dk}, and 
%an electronic health encyclopaedia maintained by the Danish Regions and 
\textit{www.min.medicin.dk} (authoritative medical health portals). % a website containing information about medical products co-funded by the Danish Regions and pharmaceutical companies.  
%We use state websites to reduce bias in the descriptions of the vaccines.
We remove stopwords and collect terms that occur in at least two different descriptions of each vaccine. 
 We treat each term as a query (i.e. we use only single term queries) and we submit it to Google Trends using Denmark as the geographical region and with the time period set to January 2011 - September 2015   
%We select The specific time period was selected since Google Trends has improved their localization method from 1 January 2011 and onwards. 
(only limited coverage of Denmark is available prior to 2011). %Because of the limited coverage we only submit single term queries to improve the retrieval. %Additionally, because  of the limited coverage of Google Trends in Denmark only one word queries were 
%submitted to increase retrieval. 
Only 58 out of 85 queries had enough coverage in Google Trends to return a result. We use these 58 queries for our predictions (shown in Table \ref{tab:queries}).

\noindent\textbf{Training.} We use as training data all data which is available prior to the data point being predicted. 
Hence if we are predicting the vaccination uptake in February 2014 we train on data from January 2011 -- January 2014. 
All models are refitted for each time step. We use monthly time steps. 
To allow for inference of seasonality, the level-0 models are initialized with 24 months of available data (January 2011 -- December 2012) as training data. 
For the level-1 models we start by using 12 months of data (January 2013 -- December 2013). 
We evaluate our predictions using the %same evaluation metrics as in \cite{yang2015accurate} namely: 
root mean squared error (RMSE), %. %, mean absolute percentage error (MAPE), correlation and correlation of the increment. 
%Let $\hat{p}$ be the prediction, $p$ the real signal and $N$ the number of items in the data set. Then:
%%\begin{equation}
%$\text{RMSE}(\hat{p}, p) = \sqrt{\frac{1}{N} \sum^N_{t=1} (\hat{p}_t - p_t)^2 }$.
%%\end{equation}
%%\begin{equation}
%%\text{MAE}(\hat{p}, p) = \frac{1}{N} \sum^N_{t=1} | \hat{p}_t - p_t |
%%\end{equation}
%%\begin{equation}
%%\text{MAPE}(\hat{p}, p) = \frac{1}{N} \sum^N_{t=1} \frac{| \hat{p}_t - p_t |}{p_t} 
%%\end{equation}
%%\begin{equation}
%%\text{correlation}(\hat{p}, p) = \frac{\sum^N_{t=1} (\hat{p}_t - \mu_{\hat{p}}) (p_t - \mu_p)}{N \sigma_{\hat{p}} \sigma_p},
%%\end{equation}
%%\noindent where $\sigma$ is the standard deviation and $\mu$ is the mean.
%RMSE 
which penalizes large errors more than small. %All RMSE are for the period January 2014 -- September 2015.%, while for MAE all errors are treated equally.

%\subsection{Tuning}

Our prediction methods are fitted using R packages with default settings at all times, except for the starting point for HW, where we manually select a starting point 
of the optimization if it cannot be completed with the default value. The AR model is trained using 12 autoregressive terms to capture seasonal variations.
%the \verb|forecasting| R package. 
%For ETS we use the \verb|ets| function, 
%which selects the ETS method with the lowest Akaike's Information Criterion\cite{hyndman2007automatic}. 
%For AR we use the \verb|Arima| function, which uses 12 autoregressive terms are used and sets all other terms to zero. 
%For ARIMA we use the \verb|auto.arima| function, which selects the model order with the lowest Akaike's Information Criterion \cite{hyndman2007automatic}. 
%For HW we use the \verb|HoltWinters| function. We use default settings at all times, except for the starting point for HW, where we manually select a starting point 
%of the optimization if it cannot be completed with the default value.
%** NDH: I have changed it back to subsets, since that is the word used in the description of bagging and weighted majority **
For bagging and weighted majority we use as many subsets as there are queries, each subset contains 10 randomly sampled queries. 
For the weighted majority we use $\eta=5$ and $\epsilon=2$ for all experiments.
%For SVM regression we use the \verb|svm| function for the R package \verb|e1071|. 

\noindent \textbf{Results.} Table \ref{tab:individual} shows the results when predicting vaccination uptake using either clinical or web data only (with the methods presented in Section 2).
 ``Naive'' refers to our naive baseline $\hat{E}(t) = E(t-1)$. 
 %We only show results for the liner model fitted with ordinary least squares since it outperformed LASSO regularization.   
 %NIELS TODO: We see that... which is overall better: clinical or web? naive or level-0 model?
Our methods outperform the naive baseline except for the HPV vaccines. This might be due to an intense debate in Denmark regarding the safety of this particular vaccine. Such a debate is likely to boost query frequencies but not necessarily vaccination uptake (the fact that many more people talk about HPV does not mean that many more HPV vaccines are given). We see that methods using clinical data outperform the methods using web data for the majority of the vaccines. But interestingly this difference is not very big and for the vaccines DiTeKiPol-3 and DiTeKiPol-4 the methods based on web data perform best. DiTeKiPol-4 is especially interesting since a shortage in 2013 resulted in unusual vaccination behaviour for a few months. When making predictions from web data our two new approaches (bagging and WM) perform best for 9 of the 13 vaccines.

%the four methods using clinical data we see that HW scores the lowest RMSE for most vaccines. For the methods using only web data the method that scores the lowest RMSE for most vaccines is the linear model with LASSO. Though if we ignore the WM method we see that bagging is the method that score the lowest RMSE for most vaccines. If we compare the results between the methods using clinical data and web data we see that methods using clinical data in the majority of the cases outperforms the web data based methods. 
%The WM, bagging and linear model with LASSO all three performs very similarly, this expected since they all build on linear models using LASSO regularization.

Table \ref{tab:ensemble} shows the results for the ensemble predictions using clinical and web data. Except for the three HPV vaccines, the ensemble approaches outperform all other methods using only one data source. We see that when using an SVR with a Gaussian kernel as level-1 model we obtain the best results, i.e.~7/13 lowest RMSE. When comparing within the methods using an SVR with a Gaussian kernel, the HW+WM is the best performing method. The most improvements are obtained when combining predictions based on web data with either predictions from HW or AR12. 
 %Except for the HPV vaccines we see that the overall best results are archived using the ensemble approaches. 
 %When combining the web data with the two simpler forecasting methods HW and AR12 we see an overall improvement in prediction quality for the majority of the vaccines. While for the two advanced methods ETS and ARIMA only limited improvements are archived. For the web data we see that the bagging and weighted majority approaches generally outperform the linear model. For the ensemble approaches we see that in the majority of the experiments methods using the bagging or weighted majority predictions also outperform the methods using predictions from the linear model.
%NIELS TODO: We see that... which is overall better: combined clinical and web or separate clinical/or web? Which ensemble method is overall best?
  
\begin{table}
\centering
\scalebox{0.6}{
\begin{tabular}{l|r|rrr|rrrr}
\hline
 & & \multicolumn{3}{c|}{Clinical data} & \multicolumn{4}{c}{Web data} \\
\hline
             &   Naive &     HW &   AR12 &   ARIMA &     WM &      B & L &     O \\
\hline
 MMR-1       &  20.704                   & \textbf{18.149} & \textbf{18.606} & \cellcolor{blue!24} \textbf{15.574} & \textbf{16.609} & \textbf{16.597} & \textbf{16.605} & 30.387 \\
 MMR-2 (4)   &  20.582                   & \cellcolor{blue!24}\textbf{13.110} & \textbf{16.566} &  \textbf{16.284} & \textbf{15.841} & \textbf{15.635} & \textbf{15.500} & 29.288 \\
 MMR-2 (12)  &  20.637                   & \textbf{19.592} & \textbf{20.600}   &  \cellcolor{blue!24}\textbf{18.726} & 21.631 & 20.815 & \textbf{21.112} & 31.897 \\
 HPV-1       & \cellcolor{blue!24} 8.080 & 11.291          & 11.192 &   9.871 & 13.474 & 14.320  & 12.701 & 11.547 \\
 HPV-2       & \cellcolor{blue!24} 8.704 & 12.522          & 12.806 &  11.276 & 18.154 & 18.025 & 18.423 & 15.404 \\
 HPV-3       & \cellcolor{blue!24} 6.579 &  9.161          & 13.958 &   9.418 & 24.239 & 23.494 & 23.074  & 17.317 \\
 DiTeKiPol-1 &  14.091                   &  \textbf{6.700} &   \textbf{5.185} &   \cellcolor{blue!24}\textbf{5.097} &  \textbf{8.067} &  \textbf{8.058} & \textbf{8.069 }& 15.913 \\
 DiTeKiPol-2 &  17.693                   &  \cellcolor{blue!24}\textbf{7.520} &    \textbf{8.030} &   \textbf{8.064} & \textbf{10.003} &  \textbf{9.941} & \textbf{9.951} & 20.082 \\
 DiTeKiPol-3 &  17.884                   & \textbf{17.596} &  20.936 &  19.459 & \textbf{17.160}  & \textbf{17.160}  &\cellcolor{blue!24} \textbf{17.158} & 30.424 \\
 DiTeKiPol-4 &  21.676                   & 26.103          &  21.676 &  23.385 &\cellcolor{blue!24} \textbf{15.414} & \textbf{15.535} & \textbf{15.934} & 33.888 \\
 PCV-1       &  13.323                   &  \textbf{6.897}  &  \cellcolor{blue!24}\textbf{6.394} &   \textbf{6.623} &  \textbf{7.745} &  \textbf{7.797} & \textbf{7.845} & 14.014 \\
 PCV-2       &  17.533                   & \cellcolor{blue!24} \textbf{7.266}  &  \textbf{8.845} &   \textbf{8.353} &  \textbf{9.679} &  \textbf{9.796} & \textbf{9.770} & \textbf{16.027} \\
 PCV-3       &  18.405                   &  \textbf{7.877}  &  \textbf{7.781} &  \cellcolor{blue!24} \textbf{7.634} & \textbf{10.410}  & \textbf{10.364} & \textbf{10.368} & \textbf{15.582} \\
 \hline
% Average & 15.837 &	12.599  &	12.070 &	13.275	& 12.290 &	14.494 &	14.425 &	21.675 \\
%\hline
\end{tabular}}
\caption{RMSE of predictions with only clinical or web data. WM: weighted majority, B: Bagging, L: linear model w. LASSO and O: linear model w. OLS. Blue: lowest RMSE per vaccine. Bold: better than naive.}
\label{tab:individual}
\end{table}

\begin{table*}[!ht]
\centering
\adjustbox{center}{
\scalebox{0.65}{
\begin{tabular}{lrrrrrrrrrrrr}
& \multicolumn{12}{c}{\large{OLS}} \\
\hline
             &   HW+WM &   HW+B &   HW+L &   HW+O &  AR12+WM &   AR12+B &   AR12+L &   AR12+O &   ARIMA+WM &   ARIMA+B &   ARIMA+L &   ARIMA+O \\
\hline
 MMR-1       &  \textbf{15.190} & \textbf{15.476} & \cellcolor{blue!24}\textbf{15.187} & \textbf{16.842} &   \textbf{17.697} & 17.572          &    17.457          &   \textbf{18.151} &   16.296          &   16.492          &  15.968          &     17.877 \\
 MMR-2 (4)   &  \cellcolor{blue!24}\textbf{12.875} & 13.497          & 13.349          &  13.121         &  16.305          & 16.108          &    16.195          &   \textbf{16.032} &   18.872          &   16.220          &  21.085          &     \textbf{15.871} \\
 MMR-2 (12)  &  \textbf{18.082} & \textbf{17.650} & \textbf{17.541} & \cellcolor{blue!24}\textbf{17.221} &  \textbf{18.711} & \textbf{19.523} &    \textbf{18.762} &   \textbf{18.909} &   \textbf{18.469} &   19.409          &  18.961          &     \textbf{18.693} \\
 HPV-1       &  \textbf{10.552} & \textbf{10.435} & \textbf{10.960} & 11.516          &   \cellcolor{blue!24}\textbf{9.377} &  \textbf{9.690} &    \textbf{10.130} &   \textbf{10.281} &   10.348          &   10.080          &   9.992          &     10.080 \\
 HPV-2       &  12.743          & \textbf{12.923} & 14.384          & \textbf{12.191} &  \textbf{11.883} & \textbf{12.201} &    13.240          &   \textbf{10.503} &   \textbf{10.708} &   \textbf{10.655} &  \textbf{10.279} &     \cellcolor{blue!24} \textbf{9.220} \\
 HPV-3       &   \cellcolor{blue!24}\textbf{8.743} &  \textbf{8.771} & 10.231          &  9.987          &  \textbf{11.321} & \textbf{11.063} &    \textbf{12.110} &   \textbf{12.151} &    9.893          &    \textbf{9.237} &   9.818          &      9.918 \\
 DiTeKiPol-1 &   \textbf{6.416} &  \textbf{6.875} &  \textbf{6.477} &  \textbf{5.498} &   \textbf{4.835} &  \textbf{4.831} &   \cellcolor{blue!24}  \textbf{4.829} &    \textbf{5.082} &    6.072          &    5.690          &   5.625          &      5.584 \\
 DiTeKiPol-2 &   9.094          &  8.216          &  8.967          &  7.956          &   \textbf{7.686} &  \cellcolor{blue!24}\textbf{7.343} &     8.116          &    \textbf{8.019} &    9.461          &    8.989          &  15.485          &      9.057 \\
 DiTeKiPol-3 &  18.478          & 17.891          & 18.410          & \cellcolor{blue!24}\textbf{16.662} &  17.529          & 18.225          &    17.550          &   \textbf{18.439} &   17.168          &   17.227          &  \textbf{17.137} &     \textbf{19.076} \\
 DiTeKiPol-4 &  \cellcolor{blue!24}15.812          & 17.977          & 17.495          & \textbf{19.860} &  17.290          & 19.891          &    16.537          &   \textbf{19.849} &   24.391          &   45.079          &  36.403          &     24.220 \\
 PCV-1       &   7.042          &  \textbf{5.783} &  \textbf{5.716} & \cellcolor{blue!24} \textbf{5.391} &   \textbf{6.201} &  \textbf{5.950} &     \textbf{5.830} &    \textbf{5.973} &   10.785          &    \textbf{6.569} &   9.174          &      \textbf{6.169} \\
 PCV-2       &   \textbf{8.317} &  8.236          &  9.283          &  \cellcolor{blue!24}7.553          &  10.135          &  \textbf{8.670} &    10.401          &    \textbf{8.284} &    8.395          &    8.399          &   9.681          &      \textbf{8.330} \\
 PCV-3       &   \textbf{7.345} &  \textbf{7.436} &  8.199          &  \textbf{7.759} &   \textbf{6.825} &  \textbf{7.014} &     \textbf{7.108} &  \cellcolor{blue!24}  \textbf{6.736} &    7.931          &    8.364          &   8.240          &      7.670 \\

& \multicolumn{12}{c}{\large{SVR linear}} \\
\hline
             &   HW+WM &   HW+B &   HW+L &   HW+O  &   AR12+WM &   AR12+B &   AR12+L &   AR12+O &   ARIMA+WM &   ARIMA+B &   ARIMA+L &   ARIMA+O \\
\hline
 MMR-1       &  \textbf{16.541} & \textbf{15.969} & \cellcolor{blue!24}\textbf{15.478}                     & \textbf{16.905} &  17.298          & 17.252          &    17.399                             &   \textbf{18.496}                    &   19.406          &   16.793          &  16.857          &     \textbf{18.204} \\
 MMR-2 (4)   &  \textbf{12.648} & \textbf{12.981} & \cellcolor{blue!24} \textbf{12.388} & \textbf{12.952} &  16.148          & \textbf{15.373} &    16.663                             &   \textbf{15.095}                    &   \textbf{14.969} &   16.101          &  \textbf{15.419} &     \textbf{15.620} \\
 MMR-2 (12)  &  \textbf{17.906} & \textbf{18.054} & \textbf{17.872}                     & \cellcolor{blue!24}\textbf{17.374} &  \textbf{19.302} & \textbf{19.020} &    \textbf{18.248}                    &   \textbf{19.075}                    &   \textbf{19.123} &   \textbf{18.193} &  19.195          &     \textbf{17.731} \\
 HPV-1       &  \textbf{10.530} & \textbf{10.649} & \textbf{10.922}                     & \textbf{11.146} &  \textbf{10.486} & \textbf{10.494} &    \textbf{10.688}                    &   \textbf{10.657}                    &   \cellcolor{blue!24}10.331          &   10.789          &  10.810          &     10.448 \\
 HPV-2       &  13.489          & \textbf{12.344} & \textbf{12.130}                     & \textbf{12.425} &  \textbf{10.147} & \textbf{10.467} &    \textbf{11.984}                    &   \textbf{11.062}                    &    \textbf{9.738} &    \textbf{9.906} &  \textbf{10.482} &     \cellcolor{blue!24} \textbf{8.727} \\
 HPV-3       &   \textbf{8.903} & \cellcolor{blue!24} \textbf{8.260} &  \textbf{8.501}                     & 10.674          &  \textbf{11.732} & \textbf{11.718} &    \textbf{13.049}                    &   \textbf{12.617}                    &    9.806          &   10.001          &   9.484          &     10.411 \\
 DiTeKiPol-1 &   6.926          &  \textbf{5.993} &  \textbf{5.739}                     &  \textbf{6.500} &   \textbf{4.740} &  \textbf{4.758} &    \cellcolor{blue!24} \textbf{4.626} &    \textbf{5.077}                    &    \textbf{5.090} &    5.354          &   5.287          &      5.507 \\
 DiTeKiPol-2 &   9.808          &  9.476          &  9.006                              &  9.100          &   8.476          &  8.563          &    10.503                             &    \cellcolor{blue!24}\textbf{7.734}                    &    9.938          &    8.480          &   9.872          &      9.591 \\
 DiTeKiPol-3 &  22.546          & 22.599          & 22.546                              & \cellcolor{blue!24}\textbf{17.433} &  21.402          & 21.925          &    21.154                             &   \textbf{19.003}                    &   22.244          &   21.319          &  22.104          &     \textbf{19.568} \\
 DiTeKiPol-4 &  16.694          & 37.909          & 17.357                              & \cellcolor{blue!24} \textbf{14.461} &  15.922          & 16.405          &    16.124                             &   \textbf{16.164}                    &   22.072          &   26.591          &  18.984          &     \textbf{17.609} \\
 PCV-1       &   7.351          &  7.186          &  \textbf{6.175}                     &  \textbf{6.198} &   \cellcolor{blue!24}\textbf{5.282} &  \textbf{6.143} &     \textbf{6.335}                    &    \textbf{5.510}                    &    6.765          &    7.413          &   6.840          &      6.830 \\
 PCV-2       &   \cellcolor{blue!24}\textbf{7.689} &  7.946          & 15.613                              &  7.955          &   9.029          &  8.879          &     9.104                             &    \textbf{8.823}                    &    8.794          &   11.662          &  14.559          &      9.024 \\
 PCV-3       &   \textbf{7.648} &  8.261          &  8.388                              &  \textbf{7.784} &   \textbf{6.904} &  \textbf{6.758} &     \textbf{6.994}                    &   \cellcolor{blue!24} \textbf{6.633} &    9.649          &    9.384          &   9.058          &      8.491 \\

& \multicolumn{12}{c}{\large{SVR Gaussian}} \\
\hline
             &   HW+WM &   HW+B &   HW+L &   HW+O  &   AR12+WM &   AR12+B &   AR12+L &   AR12+O &   ARIMA+WM &   ARIMA+B &   ARIMA+L &   ARIMA+O \\
\hline
 MMR-1       &  \cellcolor{blue!24} \textbf{14.928}& 16.694          & \textbf{16.355}                     & \textbf{16.198}                     & 17.770                             & 18.207          &    18.117          &   \textbf{16.927} &   16.703          &   17.490          &  17.569          &     17.560 \\
 MMR-2 (4)   &  14.377                             & \cellcolor{blue!24}\textbf{12.870} & 14.094                              & 13.122                              & \textbf{15.709}                    & \textbf{14.780} &    \textbf{15.024} &   \textbf{15.468} &   \textbf{14.973} &   \textbf{15.109} &  16.625          &     \textbf{15.838} \\
 MMR-2 (12)  &  \textbf{18.007}                    & \textbf{17.446} & \textbf{18.972}                     & \cellcolor{blue!24} \textbf{16.530} & \textbf{17.945}                    & \textbf{19.115} &    \textbf{18.406} &   \textbf{18.176} &   \textbf{18.385} &   \textbf{18.553} &  19.625          &     19.041 \\
 HPV-1       &  \textbf{10.748}                    & 11.606          & 11.918                              & \textbf{11.289}                     & \textbf{11.002}                    & \textbf{10.902} &    11.403          &    \textbf{9.130} &   11.664          &   10.684          &  11.505          &    \cellcolor{blue!24}  9.987 \\
 HPV-2       &  13.513                             & \textbf{11.958} & \textbf{12.304}                     & \textbf{12.097}                     & \textbf{10.789}                    & \textbf{12.376} &    \cellcolor{blue!24} 9.964          &   \textbf{12.483} &   \textbf{10.537} &   \textbf{11.249} &  11.715          &     \textbf{10.961} \\
 HPV-3       &  12.889                             & 13.784          & 14.775                              & 13.352                              & \textbf{13.016}                    & 14.521          &    15.222          &   14.374          &   12.818          &   12.172          &  \cellcolor{blue!24}12.147          &     12.682 \\
 DiTeKiPol-1 &   \textbf{5.204}                    & \cellcolor{blue!24} \textbf{5.008} &  \textbf{5.098}                     &  \textbf{5.331}                     &  5.486                             &  5.467          &     5.486          &    6.227          &    5.725          &    6.393          &   5.577          &      6.615 \\
 DiTeKiPol-2 &   7.927                             &  8.017          &  8.172                              &  9.661                              & \cellcolor{blue!24} \textbf{7.149} &  \textbf{7.443} &     \textbf{7.818} &    8.078          &    9.009          &    9.870          &   9.848          &      8.721 \\
 DiTeKiPol-3 &  \textbf{16.639}                    & \textbf{16.448} & \cellcolor{blue!24} \textbf{16.433} & \textbf{17.275}                     & 18.650                             & 18.442          &    19.009          &   \textbf{18.355} &   18.380          &   17.545          &  18.298          &     \textbf{18.962} \\
 DiTeKiPol-4 &  15.616                             & \cellcolor{blue!24}\textbf{14.877} & \textbf{15.246}                     & \textbf{16.543}                     & 16.865                             & 16.038          &    16.687          &   \textbf{15.653} &   15.938          &   15.647          &  \textbf{15.923} &     \textbf{15.932} \\
 PCV-1       &   \cellcolor{blue!24}\textbf{5.256} &  \textbf{5.808} &  \textbf{5.769}                     &  \textbf{5.664}                     &  \textbf{5.450}                    &  \textbf{6.358} &     \textbf{6.363} &    \textbf{6.614} &    \textbf{6.724} &    7.160          &   \textbf{6.611} &      7.091 \\
 PCV-2       &   \cellcolor{blue!24}\textbf{6.463} &  7.665          &  7.366                              &  7.470                              &  \textbf{8.811}                    &  \textbf{7.450} &     \textbf{6.952} &    9.026          &    8.672          &    9.062          &   8.991          &      9.148 \\
 PCV-3       &   \cellcolor{blue!24}\textbf{7.121}                    &  \textbf{7.665} &  8.396                              &  \textbf{7.798}                     &  9.022                             &  9.556          &    \textbf{10.008} &    7.871          &    8.616          &    8.148          &   9.527          &      8.176 \\
\hline
\end{tabular}}}
\caption{RMSE of ensemble predictions (clinical and web data). Blue: lowest RMSE per vaccine. Bold: lower RMSE than for the individual ensemble components in Table 1.}
\label{tab:ensemble}
\end{table*}

\section{Conclusions}

We presented a method that uses ensemble learning to combine clinical and web-mined time-series data to make predictions about future vaccination uptake. As clinical data we used official registries of vaccines in Denmark. As web data we used query frequencies collected from Google Trends. We created those queries by extracting terms from publicly available descriptions of the vaccines on the web. Experiments using all officially recommended children vaccines in Denmark for the period January 2011 -- September 2015 showed that for 10/13 vaccines our ensemble learning methods that combined clinical with web data for prediction outperformed predictions using either clinical or web data alone.
Though this combination yields the lowest overall error, using only web data gives predictions with an error only slightly worse than for the predictions made using only clinical data. This indicates the potential usefulness of web data, such as query frequencies, to predict vaccination uptake in countries where there is no national vaccination registry. This work complements wider efforts in tackling medical and health problems computationally with machine learning or retrieval \cite{Dragusin2011,doi:10.4161/rdis.25001}.

%
%We addressed the problem of..
%We used ensemble learning to combine x and y data to make predictions. 
%We experimented with z different forecasting models. 
%Experiments with all official children vaccines in Denmark for the period xxx showed that ...

%We have in this paper showed that by adding web data we can improve the prediction of vaccination uptake for 10/13 vaccines in the Danish children vaccination program. In addition we have showed queries generated using web descriptions are of high enough quality to improve overall prediction quality. Finally we have showed that prediction quality based on the query frequencies can be improved by using bagging and weighted majority methods.

\bibliographystyle{abbrv}
%\bibliography{bibliography} 

\begin{thebibliography}{}

\end{thebibliography}


\begin{thebibliography}{10}
\scriptsize

%\bibitem{dst}
%Levendefødte og døde på måneder.
%\newblock \url{http://www.statistikbanken.dk/}, December 2015.
%
%\bibitem{ssi_children_vaccines}
%Børnevaccinationsprogrammet.
%\newblock
%  \url{www.ssi.dk/Vaccination/Boernevaccination/Boernevaccinationsprogrammet.aspx},
%  January 2016.

%\bibitem{bishop2006pattern}
%C.~M. Bishop.
%\newblock {\em Pattern recognition and machine learning}.
%\newblock springer, 2006.

\bibitem{chan2011using}
E.~H. Chan, V.~Sahai, C.~Conrad, and J.~S. Brownstein.
\newblock Using web search query data to monitor dengue epidemics: A new model
  for neglected tropical disease surveillance.
\newblock {\em PLoS Negl Trop Dis}, 5(5):e1206, 2011.

\bibitem{chatfield2013analysis}
C.~Chatfield.
\newblock {\em The analysis of time series: An introduction}.
\newblock CRC press, 2013.

\bibitem{chunara2012social}
R.~Chunara, J.~R. Andrews, and J.~S. Brownstein.
\newblock Social and news media enable estimation of epidemiological patterns
  early in the 2010 Haitian cholera outbreak.
\newblock {\em The American Journal of Tropical Medicine and Hygiene},
  86(1):39--45, 2012.

\bibitem{gargano2013impact}
L.~M. Gargano, N.~L. Herbert, J.~E. Painter, J.~M. Sales, C.~Morfaw, K.~Rask,
  D.~Murray, R.~DiClemente, and J.~M. Hughes.
\newblock Impact of a physician recommendation and parental immunization
  attitudes on receipt or intention to receive adolescent vaccines.
\newblock {\em Human vaccines \& immunotherapeutics}, 9(12):2627--2633, 2013.

\bibitem{gargano2015influence}
L.~M. Gargano, N.~L. Underwood, J.~M. Sales, K.~Seib, C.~Morfaw, D.~Murray,
  R.~J. DiClemente, and J.~M. Hughes.
\newblock Influence of sources of information about influenza vaccine on
  parental attitudes and adolescent vaccine receipt.
\newblock {\em Human vaccines \& immunotherapeutics}, 2015.

\bibitem{ginsberg2009detecting}
J.~Ginsberg, M.~H. Mohebbi, R.~S. Patel, L.~Brammer, M.~S. Smolinski, and
  L.~Brilliant.
\newblock Detecting influenza epidemics using search engine query data.
\newblock {\em Nature}, 457(7232):1012--1014, 2009.

\bibitem{hyndman2007automatic}
R.~J. Hyndman and Y.~Khandakar.
\newblock Automatic time series for forecasting: The forecast package for R.
\newblock Technical report, Monash University, Department of Econometrics and
  Business Statistics, 2007.

\bibitem{kelly2009hpv}
B.~J. Kelly, A.~E. Leader, D.~J. Mittermaier, R.~C. Hornik, and J.~N. Cappella.
\newblock The HPV vaccine and the media: How has the topic been covered and
  what are the effects on knowledge about the virus and cervical cancer?
\newblock {\em Patient education and counseling}, 77(2):308--313, 2009.

\bibitem{lazer2014parable}
D.~Lazer, R.~Kennedy, G.~King, and A.~Vespignani.
\newblock The parable of Google flu: Traps in big data analysis.
\newblock {\em Science}, 343(14 March), 2014.

\bibitem{littlestone1994weighted}
N.~Littlestone and M.~K. Warmuth.
\newblock The weighted majority algorithm.
\newblock {\em Information and computation}, 108(2):212--261, 1994.

\bibitem{ma2006influenza}
K.~Ma, W.~Schaffner, C.~Colmenares, J.~Howser, J.~Jones, and K.~Poehling.
\newblock Influenza vaccinations of young children increased with media
  coverage in 2003.
\newblock {\em Pediatrics}, 117(2):e157--e163, 2006.

%\bibitem{mason2000impact}
%B.~Mason and P.~Donnelly.
%\newblock Impact of a local newspaper campaign on the uptake of the measles
%  mumps and rubella vaccine.
%\newblock {\em Journal of epidemiology and community health}, 54(6):473, 2000.

\bibitem{mciver2014wikipedia}
D.~J. McIver and J.~S. Brownstein.
\newblock Wikipedia usage estimates prevalence of influenza-like illness in the
  United States in near real-time.
\newblock {\em PLoS Comput Biol}, 10(4):e1003581, 2014.

\bibitem{paul2014twitter}
M.~J. Paul, M.~Dredze, and D.~Broniatowski.
\newblock Twitter improves influenza forecasting.
\newblock {\em PLoS currents}, 6, 2014.

\bibitem{polgreen2008using}
P.~M. Polgreen, Y.~Chen, D.~M. Pennock, F.~D. Nelson, and R.~A. Weinstein.
\newblock Using internet searches for influenza surveillance.
\newblock {\em Clinical infectious diseases}, 47(11):1443--1448, 2008.

\bibitem{santillana2015combining}
M.~Santillana, A.~T. Nguyen, M.~Dredze, M.~J. Paul, E.~O. Nsoesie, and J.~S.
  Brownstein.
\newblock Combining search, social media, and traditional data sources to
  improve influenza surveillance.
\newblock {\em PLoS Comput Biol}, 11(10):e1004513, 2015.

\bibitem{santillana2014using}
M.~Santillana, E.~O. Nsoesie, S.~R. Mekaru, D.~Scales, and J.~S. Brownstein.
\newblock Using clinicians’ search query data to monitor influenza epidemics.
\newblock {\em Clinical Infectious Diseases}, 59(10):1446--1450, 2014.

\bibitem{smith2008media}
M.~J. Smith, S.~S. Ellenberg, L.~M. Bell, and D.~M. Rubin.
\newblock Media coverage of the measles-mumps-rubella vaccine and autism
  controversy and its relationship to MMR immunization rates in the United
  States.
\newblock {\em Pediatrics}, 121(4):e836--e843, 2008.

\bibitem{yang2015accurate}
S.~Yang, M.~Santillana, and S.~Kou.
\newblock Accurate estimation of influenza epidemics using Google search data
  via ARGO.
\newblock {\em Proceedings of the National Academy of Sciences},
  112(47):14473--14478, 2015.

\bibitem{yuan2013monitoring}
Q.~Yuan, E.~O. Nsoesie, B.~Lv, G.~Peng, R.~Chunara, and J.~S. Brownstein.
\newblock Monitoring influenza epidemics in China with search query from Baidu.
\newblock {\em PloS one}, 8(5):e64323, 2013.



\bibitem{Dragusin2011}
R.~Dragusin, P.~Petcu, C.~Lioma, B.~Larsen, H.~J{\o}rgensen and O.~Winther.
\newblock Rare Disease Diagnosis as an Information Retrieval Task.
\newblock {\em ICTIR},
	356--359, 2011.

\bibitem{doi:10.4161/rdis.25001}
R.~Dragusin, P.~Petcu, C.~Lioma, B.~Larsen, H.~L.~J{\o}rgensen, I.~J.~Cox, L.~K.~Hansen P.~Ingwersen and O.~Winther.
\newblock Specialized tools are needed when searching the web for rare disease diagnoses.
\newblock {\em Rare Diseases}, 1(1):e25001, 2013.

\end{thebibliography}

\end{document}